# A Two-layer Architecture of Mobile Sinks and Static Sensors

Natarajan Meghanathan
Department of Computer Science
Jackson State University
Jackson, MS 39217

Gordon W. Skelton
Department of Computer Engineering
Jackson State University
Jackson, MS 39217

## Abstract

We propose a two-layer mobile sink and static sensor network (MSSSN) architecture for large scale wireless sensor networks. The top layer is a mobile ad hoc network of resource-rich sink nodes while the bottom layer is a network of static resource-constrained sensor nodes. The MSSSN architecture can be implemented at a lower cost with the currently available IEEE 802.11 devices that only use a single half-duplex transceiver. Each sink node is assigned a particular region to monitor and collect data. A sink node moves to the vicinity of the sensor nodes (within a few hops) to collect data. The collected data is exchanged with peer mobile sinks. Thus, the MSSSN architecture provides scalability, extends sensor lifetime by letting them operate with limited transmission range and provides connectivity between isolated regions of sensor nodes. In order to provide fault tolerance, more than one mobile sink could be collecting data from a given region or a mobile sink could collect data from more than one region. In the later half of the paper, we discuss several open research issues that need to be addressed while implementing the MSSSN architecture.

## 1. Introduction

Recent advances in electronics, embedded microprocessors, micro-fabrication and integration (example, the micro-electromechanical system technology, or MEMS) have made possible the development of low-cost, low-power and miniature sensing devices [2]. The availability of such sensors combined with the maturity of wireless network communications has lead to a new generation of massive-scale sensor networks suitable for a range of commercial and military applications. A wireless sensor network (WSN) is a distributed system of smart sensor nodes interconnected by a wireless communication network. Each sensor node is equipped with one or more sensing devices to monitor the ambient environment and collect data. The sensor node is also equipped with a processor to process the collected data and communication hardware to exchange data with other local sensor nodes within its radio range. Data collected at the sensor nodes (source nodes) is propagated to control centers called sinks where the information is required. The self-organizing ability of WSNs permits one to access data from dangerous and hostile environments which otherwise would not be possible.

Over the past few years, WSNs have been reported to be used for several applications, some of them are: Habitat monitoring [22], Environmental control [4], Vehicle tracking [40], etc. This is just the tip of the iceberg. Some of the potential applications for military and home-land security include: border patrol, power supply and nuclear plant protection, battle field surveillance, enemy troop and equipment tracking, potential threat and attack detection, etc [2]. Applications for health may include: monitoring patient's physiological parameters, doctor tracking, remote location treatment, controlled drug administration [2], rapid assistance in accident and disaster sites. Some of the environmental applications include: early bush fire detection, natural disaster early warning, weather data gathering, space exploration, wild life tracking, soil humidity and erosion monitoring and marine life monitoring. Home and commercial applications include integrated home appliance control, home security, smart transportation, inventory management and industrial process control.

With all the opportunities and promises, WSNs possess their own set of resource constraints like limited on-board battery power, network communication bandwidth, processing power, memory capacity and others [7]. Traditionally, the sinks have been static and the data collected is disseminated to the sinks using sensor-to-sensor multi-hop data propagation. To facilitate this, a routing protocol that would involve significant energy consumption needs to be executed by the energy-constrained sensors. Most of the time, data flow is from the sensors to the sinks. Sensor nodes spend lot of energy in coordinating and transmitting data through multi-hop paths to reach the sink. Nodes near the sink fail relatively earlier due to repeated relaying of data from nodes that are farther away. Sensor nodes are often not rechargeable and redeployment may not be always feasible. In such scenarios, it is essential that the lifetime of the sensors is maximized. There are several definitions for the lifetime of a sensor network. Some of them include: the first time, a region is not covered by a certain number of sensors; the time at which certain % of the sensors nodes have run out of battery power; the first time the network is partitioned.

Recently, a new category of important sensor network applications have emerged in which sinks lower the burden of the sensor nodes by being mobile [5]. Some of the applications are: battle field surveillance, wild life monitoring, locating parking spots, mobile hotspot tracking and pollution control. We propose a two-layer wireless sensor network architecture called Mobile Sinks and Static Sensor Network (MSSSN) comprising of the sensor network as the lower layer, a mobile ad hoc network of sinks equipped with wireless cameras as the top layer. The idea is mobile sinks have significant and replenishable energy resources, can move inside the area of deployment of the sensor network and collect data from the vicinity of the sensors. The MSSSN could have multiple sink nodes, each of which is assigned a certain region of the network to monitor and collect data from the sensors in that region. The mobile sinks exchange the collected data among themselves and coordinate the functioning of the entire MSSSN.

The rest of the paper is organized as follows. Section 2 reviews the earlier work with regards to employing mobile sinks in sensor networks. In Section 3, we describe our proposed MSSSN architecture and discuss its potential advantages. In Section 4, we describe open research problems associated with the proposed MSSSN architecture. Section 5 concludes the paper.

## 2. Related Work

The issue of having mobile sinks to collect data from static sensors has recently got the attention of many researchers. The strategy of deploying multiple mobile sinks in sensor networks have been viewed so far as "necessary evil". Data dissemination protocols like Directed Diffusion [17], Declarative Routing Protocol [6] and GRAB [38], suggest that each mobile sink should continuously propagate its location information throughout the sensor field to enable a sensor node to send future data reports. However, such an approach of frequently updating the locations of the mobile sinks can rapidly consume the battery power of the sensors and also cause increased collisions during wireless transmissions.

The Two-Tier Data Dissemination (TTDD) approach proposed in [21] lets each source sensor node of the data to proactively construct a grid structure such that the sensor nodes at the grid points (called dissemination nodes) forward the data from the source to the sink node. The sink node within a grid, issues a query for the data and the query is routed by the sensors within the grid to the dissemination node for the grid. The query is further propagated only by the dissemination nodes and the source now responds back through the reverse path of the dissemination nodes. Considerable overhead would be involved in establishing the grid structure for each source sensor node. The dissemination nodes at the grid points are bound to run out of battery power quickly. A variant of TTDD called the Energy Efficient Data Dissemination (EEDD) approach [41] divides the entire sensor field into virtual grids of size $R_{trans}/2\sqrt{2}$. Each grid has a grid head, most likely to be the node with the highest energy among the nodes in the grid. The grid heads are responsible for forwarding the data from the source node to the sink. The grid heads have to be frequently changed in order to maintain fairness for each sensor node. As a result, more latency will be incurred in propagating the data from a source to the sink.

In [20], the authors propose to explicitly construct a multicast tree rooted at the data source. A mobile sink associates itself with a fixed sensor node (called the access node), which acts as its proxy in the multicast tree. The proxy node is normally the node closest to the sink or the node with the maximum energy in the nearby neighborhood. In the latter case, the multi-hop path between the sink and its proxy might have to be frequently updated as the sink moves. When the sink moves far away form its proxy, a new proxy has to be selected. The method is not scalable as it requires the construction of an explicit multicast tree rooted at each sensor node that becomes a data source. The tree will have one proxy node for every sink in the network. With geographically distributed sink nodes in a large sensor network, the multicast tree will include many sensor nodes in order to span all the proxy nodes.

The Sensor Information Networking Architecture (SINA) [30] lets the mobile sink to issue a query to a particular, dedicated sensor node called the query resolver. The query resolver searches for the reply to the query either in its local cache or by interacting with the peer sensor nodes. When the reply becomes available, the resolver node forwards the reply to the mobile sink if the latter is in the neighborhood. Otherwise, the reply is forwarded through progressive footprint chaining – a sequence of logical links established from the resolver to the mobile sink as the latter moves away from the former after placing the query. The complexity of the different functionalities to be implemented at the resolver node (which is also just another sensor node) is quite high and the latency involved in transferring the data from the resolver node to the mobile sink through the sequence of logical links will also be very high.

All the previous work discussed so far consider sink mobility as something that has been imposed by the application on the sensor network. The idea of voluntarily introducing sink mobility for effective and energy-efficient data collection was explored for the first time very recently in [5], where the authors propose different sink mobility models for effective data collection. They propose purely random walk, biased random walk and deterministic walking models. Under the purely random walk model, the mobile sink moves chaotically towards all directions at varying speeds. Three models have been proposed for biased random walk: (i) the sink node has been assigned some predefined areas and the node performs random transitions from one area to another depending on their connectivity (ii) the sink gives more priority in visiting less frequently visited areas and (iii) the sink gives priority in visiting areas populated with more sensor nodes. In the deterministic walk model, the mobile sink moves along a predefined trajectory with in a small area. The trajectory is a circle of length $l$, the sink is initially on the circumference of the circle and moves around this circle of radius $r = \dfrac{l}{2\pi}$. The deterministic mobility model cannot execute complex movements. Also, there would be high overhead on the part of the sensors to constantly update the multicast trees involving the sink.

## 3. MSSSN Architecture

The MSSSN has a mix of static sensor nodes and mobile sink nodes. Each sink node is assigned a particular region of sensor nodes to control and monitor. Physically, the sensors and sinks are in the same plane. We propose logical two-layer architecture: the lower sensor network layer and the upper mobile sinks layer. To the best of our knowledge, ours is the first approach that imposes a mobile ad hoc network of sinks on the top of a sensor network. Our two-layer architecture can be implemented with the currently available IEEE 802.11 [16] devices that only use a single half-duplex transceiver. We now describe each of the two layers:

### 3.1 Sensor Network Layer

This layer consists comprises the energy-constrained, battery powered sensors that collect information about the environment and pass it to the mobile sinks. The sensors are static and the battery power is non-replenishable once exhausted. Energy is the most crucial resource of the sensor nodes and hence these devices often operate at a very limited transmission range and sensing range. We consider a homogeneous network of sensors: all sensors are from the same manufacturer and have the same transmission and sensing range.

**3.2 Mobile Ad Hoc Network of Sinks**

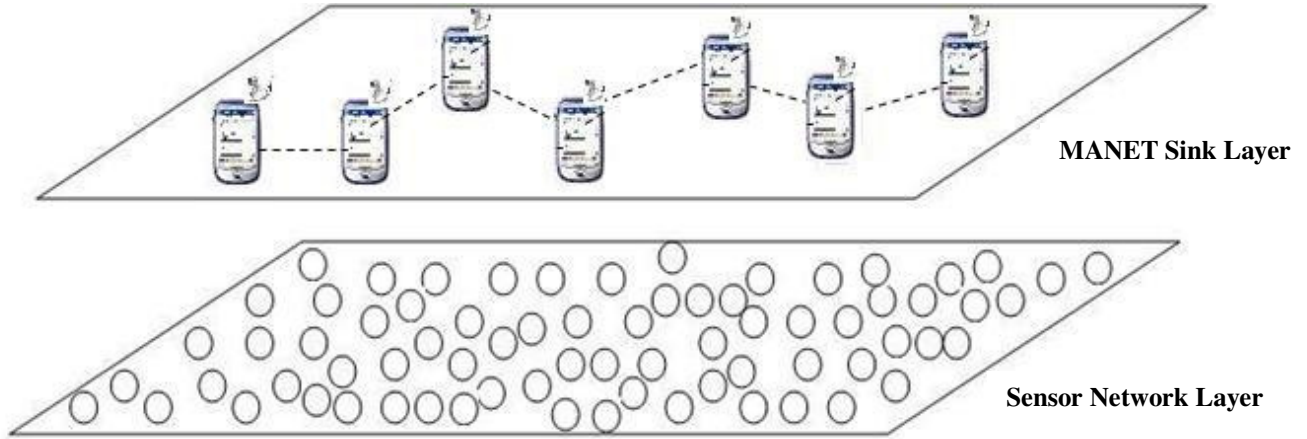

**Figure 1:** MSSSN Architecture

This layer is comprised of sinks whose main characteristic is mobility. We assume the sinks (for example, Personal Digital Assistants – PDAs) are relatively less energy-constrained (compared to the sensors in the WSN) and their main purpose is to go to the vicinity of the sensors and collect data at a reduced cost of transmission and reception energy for data propagation. The transmission range of a mobile sink would be 20-30 times to that of a sensor node. In a huge network field, a single mobile sink cannot effectively and efficiently cover the entire the sensor network. Hence, we deploy multiple mobile sinks, each assigned to cover a certain region of the network. We assume the mobile sinks are GPS (Global Positioning System) [19] enabled and hence when deployed over the MSSSN, each mobile sink will delineate the region of the sensor network it is supposed to monitor. A mobile sink collects the data from the region assigned to it, processes the data and periodically shares an aggregate of the collected data with its peers. The sink will handle localization, addressing, resource allocation and time synchronization for the sensor nodes in its assigned region. The mobility of the sinks in the sensor network field will be facilitated through a range of techniques: from simple hand-carrying to as far as automated vehicles. Integration of sink mobility with robots will be a future project. The sinks self-organize to form a mobile ad hoc network (MANET) among themselves and the communication protocols applicable in a typical MANET are applicable for this layer.

We will let each sink to be equipped with a wireless camera that can take pictures (of different zoom levels) of the region monitored by the sink. The purpose of the camera is to support the credibility of the collected data with some snapshots of the associated sensed area. The sink triggers the camera to take the appropriate picture. The data along with the snapshot is sent to the remote user of the sensor network. The remote user can also send some queries to a mobile sink. A mobile sink on receiving the query checks whether it is the appropriate sink node to respond to the query. If so, it collects the required data from its region and replies to the user. Otherwise, the mobile sink determines the appropriate sink node to answer the query. It does this by broadcasting the query to all the mobile sinks. This would be similar to the route discovery process in MANETs, except that the destination mobile sink node is not known before route discovery. The appropriate mobile sink node will then collect the data from its assigned region and if required, will co-ordinate with other mobile sink nodes. An appropriate reply is then sent back on the path that was traversed by the query packet. The mobile sink that originated the query will then receive the reply and forward it to the remote user.

### 3.3 Multi-Channel MAC Protocol

IEEE 802.11 standard for Wireless local area networks (WLANs) [16] supports multiple channels (14 channels) for use at the physical layer. The channels are 5 MHz apart in frequency. However, only 3 channels (channels 1, 6 and 11) are used in current implementations because for the channels to be totally non-overlapping, the frequency spacing must be at least 30MHz. With multiple channels, one can obtain a higher network throughput than using one channel, as multiple transmissions can occur without any interference. Unfortunately, the IEEE 802.11 Medium Access Control (MAC) Distributed Coordinate Function (DCF) protocol is designed to use only a single channel.

To use multiple channels for improving throughput, several MAC protocols like the Dual Busy Tone Multiple Access [9], Hop Reservation Multiple Access [33], Receiver Initiated Channel Hopping with Dual Polling [35], Dynamic Channel Assignment (DCA) protocol [36] and multi-channel MAC (MMAC) protocol [31] have been proposed in the literature. All of these protocols except the MMAC protocol require multiple transceivers per host and when used with the current IEEE 802.11 devices, equipped with only one half-duplex transceiver, these protocols face the multi-channel hidden terminal problem [31]. However, the MMAC protocol requires only one transceiver per host and also solves the multi-channel hidden terminal problem. With MMAC, packets transferred on two different channels do not interfere with each other.

A brief description of the assumptions and the principle of MMAC protocol are as follows: All channels have the same bandwidth. Hosts have prior knowledge of the number of channels available. As a host has only one half-duplex transceiver, the host can listen (i.e., carrier sense) or transmit on only one channel at a time. A host can switch channels dynamically with the time to switch a channel being 224μsec [16]. Clocks across all nodes are assumed to be synchronized to facilitate the beacon interval at each node to begin at the same time. At the beginning of each beacon interval, the ATIM window, each node listens onto a common channel to negotiate the channels. After the ATIM window, a node switches to its agreed channel and exchanges data on that channel for the remaining duration of the beacon interval.

### 3.4 Advantages of the MSSSN Architecture

Following are some of the advantages we see in using the proposed two-layer network architecture for the wireless sensor networks.

- **Low operational cost** – With MSSSN, we can handle sparse and disconnected networks at lower operational cost. The entire wireless sensor network need not be connected. In each region, it is sufficient for the sensors to be reachable with the mobile sink assigned to that region. A mobile sink can move into regions with fewer sensor devices and collect data by being in close proximity with such devices. Also, mobile sinks can navigate through or bypass around obstacles that block the data propagation path involving sensors alone. The mobile sinks can then co-ordinate among themselves and collect data about other regions.
- **Increased throughput:** The sensors can operate at the lowest transmission range required to just reach the mobile sinks and hence the collisions at the link level could be reduced. Also, as data propagates through fewer hops all the way from the sensor to the application user across the Internet, the probability of packet drops due to transmission error could be reduced. Hence, the network throughput could be increased.
- **Scalability and Reduced Energy Consumption:** The twin objectives of the two-layer architecture are to achieve scalability and to maximize network lifetime. Sensor networks normally employ hundreds to thousands of nodes and MSSSN supports a scalable architecture without any need for maintaining global information at the sensors. The sensor nodes are involved in multi-hop data propagation only for data originating within a narrow region and not for the entire sensor network field. Also, sensors do not need to use a larger transmission power for data packets addressed to the sink nodes. Sink nodes could be contacted with the same transmission power used to contact a neighboring sensor node. These two factors help to reduce the energy consumption at the sensors.
- **Fault Tolerance:** The carrier housing the mobile sinks could be equipped with unused, fully-battery powered sensor nodes that will be deployed in regions devoid of the required number of sensors to maintain network connectivity. In case, a mobile sink fails, the application user monitoring the network from remote can instruct a neighboring mobile sink to take control of the region devoid of mobile sink.
- **Increased data fidelity** – Communication among the mobile sinks could be protected using standard secure routing protocols for wireless networks. The number of sink nodes would be manageable and there will not be any scalability problem to employ the secure routing protocols in the MANET layer. Since, communication in the WSN layer is only for short-range, limited number of hops, data may not propagate through potentially compromised sensor nodes that forward data to an adversary.

## 4 Open Research Problems

The proposed MSSSN architecture generates a lot of open research problems that need to be addressed. In this section, we discuss the open research issues currently being addressed by us and those that will be addressed in the near future.

### 4.1 Sensor-Sink and Sink-Sink Communication

We are currently investigating two possible options for sensor-sink and sink-sink communication. One option is to have each sink equipped with a sensor node that acts as a proxy for the sink and communicates with the underlying sensor network. The sink replaces its proxy sensor node as it runs out of energy. We assume the sink has a pool of backup sensor nodes which are used as proxy nodes and also as replacement nodes for the battery-drained sensors in the field. The sensors in the field communicate with the proxy sensor, which is connected to the mobile sink through an Ethernet cable. The mobile sinks self-organize themselves to form an ad hoc network and communicate using MANET routing protocols and broadcasting strategies.

The other option (instead of using a proxy sensor) is to use a multi-channel MAC protocol for sink-sink and sink-sensor communication. We will use the MMAC protocol for sensor-sink (channel 1), sensor-sensor (channel 6) and sink-sink (channel 11) communication. The main objective would be to maximize the throughput and at the same time minimize the interference between the sink-sink communication spanning long distance and the sensor-sensor communication spanning over short distances. We are investigating both these options and are also looking at hardware options to best realize this architecture. Note that the mobile sinks may sometimes need to collect data only on-demand (i.e., only when required). We will survey the existing approaches and also new approaches for using parallel channels to wake up the sensor nodes through the mobile sink.

### 4.2 Broadcasting Strategies and Routing Protocols for Sink-Sink Communication

The type of communication involved in the mobile sink ad hoc network would be a multi-hop communication between any two mobile sinks that want to co-ordinate among themselves using a particular unicast routing protocol or a broadcast query originating from a particular sink asking for certain data from one or more sinks. We will investigate the use of simple flooding, probability-based, area-based and location-based approaches for efficient broadcasting [24]. The unicast routing protocols studied would be the minimum-hop based Dynamic Source Routing (DSR) [18], Ad Hoc On-Demand Distance Vector (AODV) [26] routing protocols and the stability-based Associativty-based Routing (ABR) [34], Flow-oriented Routing (FORP) [32] and Route-lifetime Assessment Based Routing (RABR) [1] protocols. For different practical scenarios, we will identify the appropriate broadcast strategies and unicast MANET routing protocols that would give the best performance for the MSSSN. The performance metrics studied would be number of route transitions, delay and energy consumption.

### 4.3 Tracking a Mobile Hotspot through Collaboration between Mobile Sinks

It is often the case that the data collected or recorded in one region of the sensor network need to be made available to the other regions of the network. If it were a flat sensor network topology, it would be too much of an overhead for sensor nodes at two extremes of the network to often collaborate. With a hierarchical architecture in MSSSN, a mobile sink will collect information from the sensor nodes in its region, analyze and aggregate the data and then exchange the data with peer mobile sinks. One useful application would be to track a mobile hotspot. A mobile sink tracking a hotspot in its region alerts the neighboring peer sinks about the possible entry of the hotspot into their region and also sends them the data it had already collected about the hotspot. We need to identify the appropriate mobile ad hoc network routing protocol for sink-sink coordination and the sensing data dissemination protocol for sink-sensor coordination to effectively and efficiently track a mobile hotspot.

### 4.4 Developing Sink Mobility Models

One issue that needs to be addressed very seriously in MSSSN is the mobility model for the sinks. To quickly receive the data from the sensor nodes and minimize the energy consumption at the sensors, the number of intermediate sensor nodes for the communication between the mobile sink and the source sensor node for the data should be as low as possible. Hence, the mobile sink should roam around in the vicinity of sensor nodes that collect useful data. On the other hand, the mobile sink cannot just directly communicate with each sensor. We will develop a data-driven mobility model rather than adopt a pure random-walk model or a deterministic movement model. The data-driven mobility model will let the mobile sink communicate with the source sensor node through the minimum number of hops and also move the sink node towards other source sensor nodes for data collection within a limited latency. To determine the next location of movement, we need to be able to predict the "usefulness" of the data that would be collected from the location, the "usefulness" of the data that was collected previously from that location and the time elapsed since data was last collected from that location. The "usefulness" of data at any time will be application specific.

### 4.5 Determination of Multicast Steiner Trees in a Region

As the mobile sink keeps moving in a region, the communication structure between the sink and the sensor nodes in the region keeps changing with time. We are considering two possible approaches: Use the mobile sink (i.e., the proxy sensor node at the sink) itself as the root and determine a stable multicast Steiner tree spanning over multiple sensor nodes; Select a sensor node in the assigned region as access node, establish a multicast tree rooted at the access node and let the mobile sink communicate to the access node through a unicast multi-hop path. In either case, we need to keep updating the multicast tree over a period of time. We will evaluate the pros and cons of the two approaches and determine the suitable approach for appropriate conditions. An algorithm to determine the sequence of stable multicast Steiner trees in mobile ad hoc networks has been proposed in [23]. Active work is currently in progress to develop a routing protocol that constructs a stable multicast tree in a distributed fashion.

### 4.6 Difference with Cluster-head Approach

Our hierarchical approach in MSSSN is drastically different from the cluster-head approach used in some of the well-known protocols like LEACH [14]. Ours is a heterogeneous system of static sensor nodes and mobile sinks. The sinks in our case, are assumed to have high memory capacity plus processing power and are not much energy constrained. Once assigned, the mobile sink for a region is never changed unless the sink fails. In LEACH, the role of cluster head has to be frequently rotated among all sensor nodes. The number of nodes in LEACH clusters is such that the number of hops between any two nodes in a cluster is at most 2. We do not have such restrictions; nevertheless will need to determine the appropriate mobile sink to sensor ratio for different application scenarios and topology fields.

### 4.7 Localization Strategies

We will investigate localization based on the raw Received Signal Strength Indicator (RSSI) [28] and quantized RSSI values [25]. If the measurements have too much of errors, then we will try the time of arrival approaches [29], which require two types of senders and receivers (to generate sound waves and radio waves) per node. We will use the mobile sinks as the anchor nodes for the tri-lateration process. The mobile sinks are assumed to be GPS enabled. During the initialization phase of the MSSSN, the mobile sinks will synchronize among themselves and arrive at a time schedule on when to broadcast a beacon signal (containing information on the location of the mobile sink) that will reach the sensor nodes in their own region and the sensor nodes in the neighboring regions. A sensor node on receiving a beacon signal will extract the location information of the sender and also estimate the distance to the sender based on the received signal strength. Once a sensor node has received three beacon signals, the node positions itself at the center of the intersections of the circles around these anchors.

### 4.8 Time Synchronization Strategies

There are two strategies for time synchronization in wireless sensor networks: sender/receiver based synchronization – the receiver of a timestamped packet synchronizes with the sender of the packet and receiver/receiver based synchronization – multiple receivers of a timestamped packet synchronize with each other, but not with the sender. We will use the mobile sink as the reference node for synchronization within its assigned region during the initialization phase of the MSSSN. The time synchronization protocols studied would be: Lightweight Time Synchronization (LTS) protocol [13], Timing-sync Protocol for Sensor Networks (TPSN) [12], Reference Broadcast synchronization (RBS) [10] and Hierarchy Referencing Time Synchronization (HRTS) [8]. LTS and TPSN are based on sender/receiver synchronization, while RBS and HRTS are based on receiver/receiver synchronization.

### 4.9 Secure Routing between Mobile Sinks

Mobile sinks are vulnerable to get compromised with potential intruders. Hence, we would also study securing the communication between the mobile sinks. In addition to the minimum-hop based and stability based routing protocols for sink-sink communication, we will also investigate the use of the following security-based routing protocols for ad hoc networks: Security-aware Ad Hoc Routing (SAR) [39], Secure Efficient Ad Hoc Distance Vector (SEAD) Routing protocol [15] and Authenticated Routing protocol for Ad hoc Networks (ARAN) [27]. We will conduct a performance comparison analysis of these three secure routing protocols which has not been done so far in the literature.

## 5   Conclusions

The high-level contribution of this paper is the proposal of a scalable two-layer mobile sink and static sensor network (MSSSN) architecture for wireless sensor networks. The top layer includes a mobile ad hoc network of sink nodes and the bottom layer includes the static sensor nodes. The architecture could connect isolated regions of sensor nodes using multiple mobile sink nodes, each collecting data from a certain region and exchanging the collected data with peer sink nodes. The architecture could be realized with existing IEEE 802.11 devices that use only a single half-duplex transceiver. The architecture leads to lots of open research issues like developing sink mobility models, determining multicast trees connecting the mobile sinks with static sensors, tracking a hotspot through collaboration between multiple mobile sinks, etc. We are currently analyzing these research issues in a simulator environment. In the near future, we are planning to implement the MSSSN architecture with a mobile ad hoc network of Personal Digital Assistants (PDAs) and the Mica motes [37] as static sensor nodes. We are also considering using the Freescale [11] and Archrock [3] development and evaluation kits for static layer of sensor nodes.